\documentstyle[epsfig,11pt,hrd_pasp,twoside]{article}

\begin{document}

\title{On some Uncertainties in Evolutionary Synthesis Models}

\author{M. Cervi\~no}
\affil{Max-Planck-Institut f\"ur extraterrestrische Physik, 
              Giessenbachstrasse, D-85748 Garching bei M\"unchen, Germany}
\author{D. Valls--Gabaud}
\affil{UMR CNRS 5572, Observatoire Midi-Pyr\'en\'ees, 
	      14, avenue Edouard Belin,  31400 Toulouse, France}

\begin{abstract}
Ranging from track interpolation techniques through model atmospheres 
to the stochastic nature of the IMF, there are many uncertainties
which need to be taken into
account when modelling HR diagrams or performing population synthesis,
particularly if comparison with actual data is sought. In this paper,
we highlight (1) the problem of discontinuities along evolutionary
tracks of massive stars (M$>$ 8 M$_\odot$), showing that
inconsistencies appear in the computation of the corresponding isochrones, 
and (2) the sampling fluctuations produced by the stochastic nature
of the IMF, presenting a statistical formalism to estimate the 
dispersion in any given observable of a stellar population due 
to sampling effects which bypasses the need of performing Monte Carlo
simulations.
\end{abstract}

\section{Introduction and motivation}

In the last few years, the increasingly detailed observations of stellar
populations in a wide variety of environments have provided a huge amount of
high quality data, which have been used to constrain both 
stellar evolution theory and stellar models among many other variables.

In contrast, the development of increasingly complicated evolutionary
synthesis codes has, in general, focused only on the use of updated
physical ingredients, but some of their underlying hypotheses (track 
interpolations
and isochrone computation in particular) have remained unchanged.  
In addition, the study
of sampling  errors has received comparatively little attention (see however
 Buzzoni (1989), Lan\c{c}on (these proceedings), and Cervi\~no et al. (2001b)
for a more complete review) even though it seems likely that the scatter in
the observed properties of systems with a relatively small number of stars
can be accounted for by sampling fluctuations from a given (and perhaps
universal) stellar initial mass function (IMF).  The aim of this
contribution is to highlight some oversimplifications used in
these evolutionary synthesis models which lead inevitably to larger
uncertainties in some integrated quantities. We present 
a short general overview of how
evolutionary synthesis models work (Sec. \S2), the problems raised by 
discontinuities in stellar tracks of different masses for the isochrones 
of massive stars (Sec. \S3), 
and the effects of the stochastic nature of the IMF in integrated
quantities (Sec. \S4).

\section{How evolutionary synthesis models work}

Irrespective of the actual technique used to obtain the {\it
integrated} properties of a stellar cluster 
as a function of age several steps must be followed. For illustration
we focus here on a fixed metallicity and an instantaneous burst of
star formation, and  we assume that we have a set of 
stellar tracks
discrete both in the number of stellar masses and in the 
evolutionary phases provided. As an example, 
we take here the widely used solar metallicity tracks from 
Schaller et al. (1992). The {\it
first step} is to interpolate the tracks in order to obtain a set of {\it
continuous} distributions of stellar tracks that describe the same
evolutionary sequence for a denser set of initial stellar masses.  
The {\it second step}
is to assign the age for each mass and each evolutionary sequence in the
previously computed stellar tracks.  At this stage, the isochrone can be
derived from interpolations between the computed tracks at a given age.
The next step is to populate the isochrone with the number of stars
in a given mass interval. The total number can match the number of
stars observed in a cluster, but note that the number of stars in
each mass interval is given by the Initial
Mass Function (IMF) only {\it in the asymptotic limit of an
infinite number of stars in the cluster}. As an example, 
we adopt here a power law IMF in the mass range 2--120 M$_\odot$
with a Salpeter slope.

Additional complications related to chemical evolution or star
formation history introduce other assumptions that will not be dealt
with here.  It is important to note that although this study is 
restricted to a narrow range of possibilities 
(massive stars and young clusters), entirely similar 
situations are found in  older clusters with low mass stars which 
undergo the Helium flash.

\section{Homology relations and isochrone calculations}

\subsection{Oversimplifications in the use of homology relations}

Over the limited region of parameter space where the opacity, $\kappa_0$,
and the energy production rate, $\epsilon_0$, of the star can be represented by
power laws, the internal structure of stars can be assumed to be
homologous. This is particularly useful to get, for example, 
the absolute luminosity, $L$, as a function of the mass of the star:

\begin{equation}
L(M) \;  \propto \; \epsilon_0^a \, \kappa_0^b \, \mu^c\, M^d \;
\propto \; M^d
\end{equation}

\noindent where $\mu$ is the mean molecular weight. If the variations of
$\epsilon_0$, $\kappa_0$ and $\mu$ in  the tabulated tracks are  small
enough,  interpolations in the $\log L_k - \log M$ plane (where $k$
represents an evolutionary stage) are valid.

Unfortunately, in the case of massive stars it is not always the case.
First of all, the initial mass of the star and the mean molecular
weight $\mu$ are not constant due
to the effects of stellar winds. Second, there is obviously 
a strong discontinuity between
the evolution of stars that ultimately  become Wolf-Rayet (WR) stars,
and which are placed on the left side
of the Main Sequence, and stars which become red supergiants. 
(Note that a similar discontinuity appears for low mass stars which
undergo the Helium flash).  Such a discontinuity could  be solved if the
sets of evolutionary tracks would give  additional tables corresponding to
the appropriate mass range. In the case of Helium flash stars, see for
example the solution adopted by Brocato et al. (1999). No such extra
information is currently available for this critical range in 
massive stars, hence there are unavoidable uncertainties associated
with this discontinuity. 

\subsection{Isochrone computations}

Assume however that this problem can be bypassed, somehow, and let's
go
to the next step: a linear interpolation in the $\log M - \log t_k$
plane is used. Such a relation is certainly valid at the 
Zero Age Main Sequence, but it  fails as soon as the stars evolve. 
This situation is particularly well illustrated by the calculation
of the supernovae rate (SNr). If the relation between $M$ and
the age where the star becomes a SN, $t_{SN}$, is a power law, and if the
IMF is another power law, the evolution of the SNr must be another power
law since it is a convolution of two power laws. We show in Fig.~1 the
SNr using linear and parabolic interpolations in the $\log M - \log t_{SN}$
plane.

\begin{figure}
\centering \epsfig{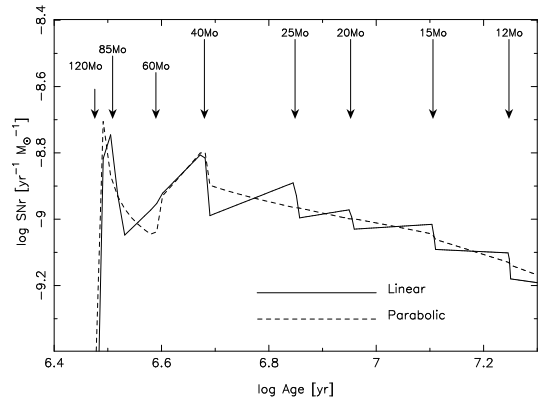} \caption{The
supernovae rate SNr  using two 
 different interpolation techniques. The solid line assumes a linear
 interpolation in the $\log M -\log t$ plane.  The short-dashed line
 corresponds to a parabolic interpolation.}
\end{figure}

While the SNr using the linear interpolation exhibits discontinuities
corresponding to the discreteness of the stellar tracks,  the parabolic
interpolation scheme presents a much smoother behaviour. 
However, independently of the 
interpolation technique, there are some wiggles  at the beginning of the
evolution of the SNr due to the particular behaviour of the lifetime of WR
stars present in the set of tracks used. However, even if the parabolic
interpolation seems to produce better  results, a correct
interpolation technique (based on physical principles) is still lacking,
and a more careful study is required on this subject.  A more detailed
assessment of this problem can be found in Cervi\~no et al. (2001a).

\section{Stochasticity of the IMF}

Besides the problems mentioned in the previous Section, let's turn to
the effect of sampling  the IMF to evaluate the dispersion due
to the discreteness of actual stellar populations.  Let us assume that 
$N_{tot}$ stars are observed, with masses distributed between $M_{low}$ and
$M_{up}$. The probability distribution function of the stellar  masses is
given by the IMF, so that the number of stars $N_i$ of a given mass
$M_i$ is a random variable with a Poissonian distribution (see
Cervi\~no et al. 2001b for details).

As pointed out by the pioneering study of Buzzoni (1989), the 
fact that $N_i$ is 
a Poisson variable makes it possible to apply a proper statistical
formalism. Let us consider for instance the integrated monochromatic
luminosity ${{L}}_{\lambda}$ of a stellar population of N$_{tot}$ ~stars of
a given age $t$. The average value of ${{L}}_{\lambda}$ is given by the
sum of the individual monochromatic luminosities $l_{\lambda,i}$ of stars
of age $t$ belonging to the $i^{th}$ mass, each weighted by its number
as given by the IMF, $w_i = N_i / N_{tot}$:

\begin{equation}
< { L}_{\lambda}> \; = \; N_{tot} \, \sum_{i} \, w_i(t) \, l_{\lambda, i}(t).
\label{eq:lum}  
\end{equation}

Up to a constant, ${ L}_\lambda (t)$ will be Poisson distributed, since the sum
of Poisson variables is also a Poisson variable, with parameter given by
the sum of the individual Poisson parameters.  Due to  random
fluctuations in the number of stars of each mass, the actual luminosity
${{L}}_\lambda$ will fluctuate around this average value with a variance
given by

\begin{equation}
\sigma^2({ L}_{\lambda})  = \; < \left
( { L}_{\lambda} - < { L}_{\lambda} > \right)^2 > 
\; = \; N_{tot} \sum_i \, w_i \, l_{\lambda, i}^2
\end{equation}

and thus the dispersion due to the discreteness of the stellar population is
obtained for any quantity that is derived from the integrated
luminosity, such as colours or the number of ionising photons. Note
however
that for {\it derived} quantities one has to take into account the
{\it covariances} between the $w_i$, weighted by their luminosities.
This can be illustrated by the correlation 
coefficient of two quantities
$\rho(x,y)$ which is defined as:

\begin{equation}
\rho(x,y) = \frac{{{cov}}(x,y)}{\sigma_x \, \sigma_y}.
\end{equation}

It obviously varies between $-1$ and $+1$, where the sign indicates the sense of the
correlation. So $\|\rho(x,y)\| = 1$ if the
quantities are completely correlated and $\rho(x,y) = 0$ if there is no
correlation.  For a given star, the luminosities in different bands are
completely correlated since they are produced by the same star.  But the
situation changes when several stars are considered simultaneously,
such as in a cluster. At some ages, the same stars will contribute
to the same bands, and hence the correlation coefficient will be
unity for these bands, but at some later stages different masses
will contribute differently and the correlation will decrease. As an
illustration, Fig.~2 show the evolution of the correlation coefficient
for some optical and NIR bands. Neglecting this evolving coefficient
may produce an underestimation of  the sampling errors. Similarly some
bands may be more affected by the nebular continuum (strongly
correlated
to the more massive stars) than others, and hence the actual
dispersion on a given magnitude or colour 
will depend on the precise correlation  coefficient, as shown on Fig.~2.

\begin{figure}
 \resizebox{\hsize}{!}{\epsfig{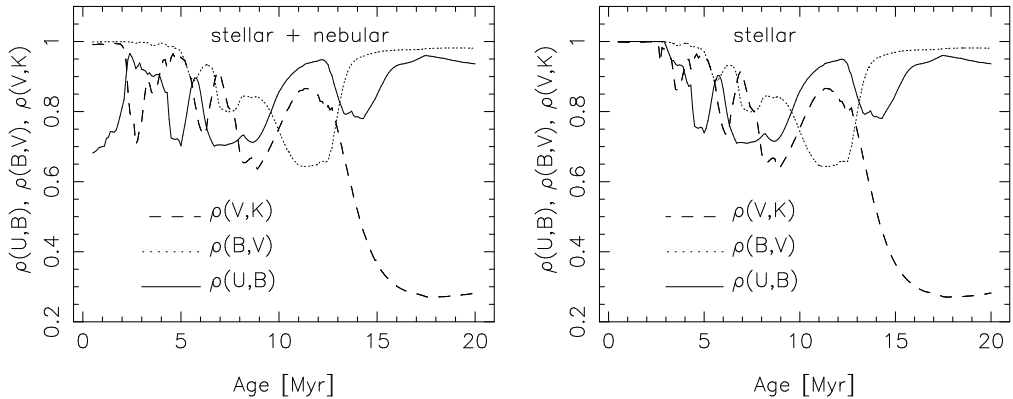}}
 \caption{{\it Correlation coefficients.} Evolution of the correlation
 coefficients for $U$ and $B$, $B$ and $V$ and $V$ and $K$ for a cluster where the
 nebular contribution is included (left panel) and for a cluster with only
 stellar contribution (right panel).}
\label{fig:app2}
\end{figure}

Using simple rules of error propagation, and taking into account
the covariance terms, we can derive {\it analytically}
the dispersion in any integrated quantity derived from the
luminosity. Fig.~3 shows the interesting case of the $V-K$ colour as
an illustration.  
We have performed Monte Carlo simulations with  1000 clusters with
$N_{tot}$=10$^3$ stars, 500 clusters with 10$^4$ stars and 100 clusters
with 10$^5$ stars. In each set we have obtained the dispersion
$\sigma_{clus}$(V--K).  The $\sigma_{clus}$ values have been divided by
$N_{tot}^{0.5}$, to obtain a normalized effective dispersion  
for each set (see Cervi\~no et al. 2001b for further details).

\begin{figure}
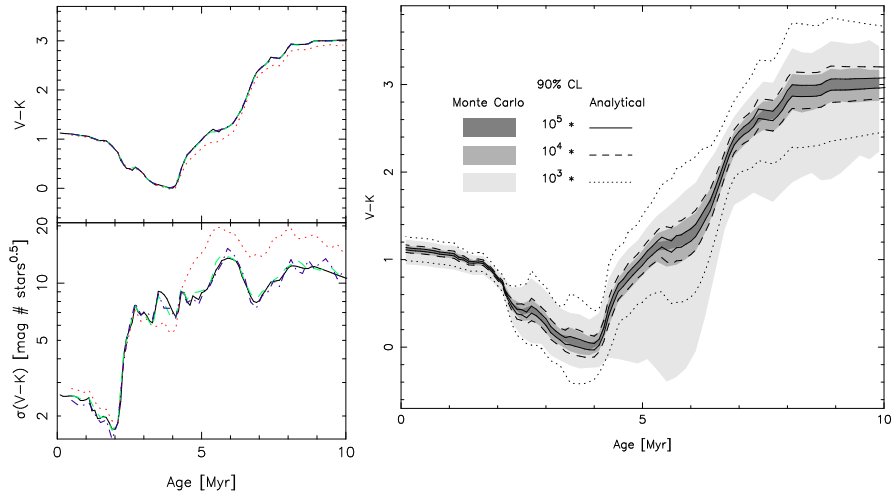

\centering {\epsfig{angle=270,width=4.6cm,file=cervino_fig3a.eps}
\epsfig{angle=270,width=7cm,file=cervino_fig3b.eps}}
\caption{{\it Comparison of analytical versus Monte Carlo dispersion
estimations}. {\it Top left:}
Evolution of the average $V-K$ color index for the analytical (solid line) and Monte
Carlo simulations (dashed/dotted).  {\it Bottom left:} Comparison of
normalised dispersions  
$\sigma(V-K)$  for  analytical and Monte Carlo estimations.{\it
Right:} Comparison of the analytical estimation of the confidence levels
(lines) with  Monte Carlo ones (shadowed areas).}
\end{figure}

The top left panel shows the mean $V-K$ colour index for  different
simulations and the analytical one (the idealised case of an infinite
number of stars).  The comparison shows that our analytical estimation
performs very well, although  there are slight deviations in 
the 10$^3$ stars cluster simulations.  The low left panel shows the comparison of the normalized
values of $\sigma_{clus}(V-K)$ with the analytical value.  It shows that
the analytical computation of the dispersion and the Monte Carlo results
are again very similar except for the 10$^3$ star clusters. The differences are
due to an additional dispersion introduced by the relation between
N$_{tot}$ and the total mass of the cluster.  We also show the 90\%
confidence levels of the different simulations computed from the Monte
Carlo simulations (shadowed areas) and analytically (lines). It is important
to stress that these confidence level bands are not {\it symmetric} with
respect to the mean value, as expected for quantities derived
from a Poisson distribution. We refer the reader to Cervi\~no et al. (2001b) for more
details.

\section{Conclusions}

Evolutionary synthesis models clearly have oversimplifications that must be
studied in much more detail if reliable predictions are to be made. 
Some of them can be solved, at least in part,  if evolutionary
tracks covering a denser grid in mass are published. The effects
of the discreteness of stellar populations in small systems must also
be considered before a comparison of model predictions to real data
is attempted.

\begin{acknowledgements}
We warmly thank the LOC for financial support to attend the meeting.
\end{acknowledgements}

\bibliographystyle{apj}

\begin{thebibliography}{}

\bibitem[Brocato et al. (1999)]{Bro99} Brocato, E., Castellani, V.,
   Raimondo, G., \& Romaniello, M. 1999, A\&AS, 136, 65
\bibitem[Buzzoni~(1989)]{Buzz89} Buzzoni, A. 1989, ApJS, 71, 871 \\
{\tt http://www.merate.mi.astro.it/$\sim$eps/home.html}
\bibitem[Cervi\~no et al. (2001a)]{Cetal01b} Cervi\~no, M., 
   G\'omez-Flechoso, M.A., Castander, F.J., Schaerer, D., Moll\'a, M., 
   Kn\"odlseder, J., \& Luridiana, V. 2001a, A\&A, 376, 422
\bibitem[Cervi\~no et al.~(2001b)]{CVGMH01} Cervi\~no, M.,
   Valls-Gabaud, D., Luridiana, V. \&  Mas-Hesse, J.M. 2001b, A\&A, (accepted)
{\tt http://www.laeff.esa.es/users/mcs/}
\bibitem[Lan\c{c}on \& Mouhcine (2000)]{LM99} Lan\c{c}on, A., \&
   Mouhcine, M. 2000, in {\it Massive Star Clusters}, ASP Conf. Series  
   Vol. 211, p. 34
\bibitem[Schaller et al. (1992)]{Schetal92} Schaller, G., Schaerer, D.,
   Meynet, G., \& Maeder, A. 1992, A\&AS, 96, 269
\end{thebibliography}
\end{document}